# Communications Convergence, Spectrum Use and Regulatory Constraints,
# Or
# Property Rights, Flexible Spectrum Use and Satellite v. Terrestrial Uses and Users

Douglas W. Webbink[1]
International Bureau
Federal Communications Commission
Washington, D.C.

9/13/2001

Introduction

As far as many consumers and businessmen and women are concerned, increasingly wireline and wireless services, including those provided by terrestrial and satellite systems, are considered to be substitutes and sometimes complements, regardless of the laws and regulations applicable to them. At the same time, many writers and even government agencies (such as the FCC) have suggested that users of the spectrum should be given more property-like rights in the use of the spectrum and at a minimum should be given much more flexibility in how they may use the spectrum. Two recent developments have important implications with respect to "convergence," spectrum property rights and flexible use of the spectrum. The first development involves several proposals to provide terrestrial wireless services within spectrum in use or planned to be used to provide satellite services. The second development is the passage of the 2000 ORBIT Act which specifically forbids the use of license auctions to select among mutually exclusive applicants to provide international or global satellite communications service. The purpose of this paper is to discuss some of the questions raised by these two events, but not necessarily to provide definitive answers or solutions.

Convergence and Spectrum Use

Communications services and technologies appear to be converging faster than laws and regulations can adjust to those changes. Consumers (including businessmen and women) often view various services and products as substitutes or complements, regardless of their regulatory status. For example, many consumers consider services supplied by TV stations, cable TV systems, DBS operators and even streaming video over the Internet to be substitutes and sometimes complements. Increasingly consumers see wireless and wireline telephones as substitutes and sometimes as complements. The same consumer reaction likely applies to FM and AM radio signals, audio signals delivered over cable TV and DBS systems, streaming audio over the internet and in the future, satellite digital audio radio service (DARS). Wireless phones, personal digital assistants and computers may also be in the process of converging. These examples also suggest that providers of services will increasingly recognize the significance of this substitutability or complementarity and may wish to provide many of these services using the same spectrum as well as wire or fiber optic technology, regardless of the regulations or laws to which they are subject.

---

[1] The views expressed are my own and should not be interpreted to be those of the FCC or any of its Commissioners or staff. I am grateful for useful suggestions from Karl Kensinger and David Sappington. Errors and opinions remain my own.

Property Rights and Flexible Spectrum Use

Over the last 40 years, a large number of authors have argued that users of the radio frequency spectrum, i.e. licensees, should be given property rights or at least more property-like rights in the use of the spectrum.[2] Such rights would include the right to purchase, sell, lease, or give away the resource, the right to combine and subdivide the resource, the right to use it as one wishes, including the right to use as little or as much of it as one wishes, the right to develop or not develop its use at a rate determined by the property right owner and the right to decide who can use the resources and who can be excluded.[3] Generally, these authors argue that only by giving licensees complete or at least more complete property rights, will those licensees face the correct economic incentives to use the spectrum efficiently, i.e. to take into account the opportunity cost of the choices they make with respect to spectrum usage, to consider the tradeoffs between expenditures on acquiring hardware, software and additional spectrum, and to make efficient decisions about investment, technological innovation and changes in the services they decide to provide.

---

[2] See, for example, Lawrence J. White, "'Propertizing" the Electromagnetic Spectrum: Why It's Important, and How to Begin," Media Law and Policy (Fall, 2000), pp. 51-75; Pablo T. Spiller and Carlo Cardilli, "Towards a Property Rights Approach to Communications Spectrum," Yale Journal on Regulation, vol. 16 (Winter 1999), pp. 53-83; Howard A. Shelanski and Peter W. Huber, "Administrative Creation of Property Rights to Radio Spectrum," Journal of Law and Economics, vol. XLI (October 1998), pp. 581-607; Glenn O. Robinson, "Spectrum Property Law 101," Journal of Law and Economics, vol. XLI (October 1998), pp. 609-625; Arthur De Vany, "Implementing a Market-Based Spectrum Policy," Journal of Law and Economics, vol. XLI (October 1998), pp. 627-646; Douglas W. Webbink, "Radio Licenses and Frequency Spectrum Use Property Rights," Communications and the Law, 9 (June 1987), pp. 3-29; Milton Mueller, "Property Rights in Radio Communications: The Key to Reform of Telecommunications Regulation," Cato Institute Policy Analysis (June 3, 1982); Louis De Alessi, "The Economics of Property Rights: A Review of the Evidence," in Research in Law and Economics, vol. 2, edited by Richard O. Zerbe, Jr., (1980), pp. 1-47; Jora R. Minasian, "Property Rights in Radiation: an Alternative Approach to Radio Frequency Allocation," Journal of Law and Economics, vol. 18 (April 1975), pp. 221-272; Arthur S. De Vany, Ross D. Eckert, Charles J. Meyers, Donald J. O'Hara and Richard C. Scott, "A Property System for Market Allocations of the Electromagnetic Spectrum: A Legal-Economic-Engineering Study," Stanford Law Review, vol. 21 (1969), pp. 1499-1566: Ronald H. Coase, "The Federal Communications Commission," Journal of Law and Economics, vol. 2, (1959), pp. 1-40.

For a different view, that suggests that granting exclusive property rights using auctions may impose a variety of costs on society, including creating incentives for oligopoly industry structures, see: Eli Noam, "Spectrum Auctions: Yesterday's Heresy, Today's Orthodoxy, Tomorrow's Anachronism. Taking the Next Step to Open Spectrum Access," The Journal of law and Economics, vol. XLI, no. 2 (October 1998), pp. 765-790. For a response to Noam, see: Thomas W. Hazlett, "Spectrum Flash Dance: Eli Noam's Proposal for "Open Access" to Radio Waves," The Journal of Law and Economics, vol. XLI, no. 2 (October 1998), pp. 805-820.

[3] According to Shelanski and Huber, "complete" property rights entail the right to hold, transfer, subdivide, or use the property in any way one sees fit and to exclude anyone. Shelanski & Huber, pp. 583-84.

In addition, a number of authors and the Federal Communications Commission (FCC) have suggested in rulemaking proceedings that licensees should be given increased flexibility in the ways in which they can use the spectrum to which they are licensed, even if they are not given full property rights in spectrum use.[4] Such flexibility would include the right to decide how much of the spectrum licensed to an individual to use and for what purpose, what technologies to use and what services to provide, etc.

Of course, even if one claims to be in favor of moving towards more property-like rights in spectrum use and increased flexibility in allowable uses of the spectrum, making such general statements does not begin to explain the details of how to implement such concepts. For example, what degree of exclusive property rights should be allowed and with how many restrictions? Similarly, how much flexibility should be allowed and what limits should be placed on that flexibility? What is the government role in defining the initial bundle of property rights granted, and to what extent should the government be able to modify that bundle of rights for incumbent users? To what extent do increasing flexibility and allowing more property-like rights make resolving interference disputes more complex rather than easier and thus increase transactions costs among affected parties? These questions are similar to questions about the extent to which land owners should have rights to use their property as they wish, and the extent to which they should be subject to overall zoning laws and other restrictions on allowable uses. In every case, a key issue is the extent to which private parties acting alone, or a government agency, should attempt to resolve these issues.

Satellite v. Terrestrial Spectrum Use

Recently, the question of whether it is possible and desirable to give more flexibility and more property-like rights to spectrum or radio licensees has come up again in the context of a number of specific proposals and proceedings involving the possible use of the same block or section of spectrum by satellite and by terrestrial wireless systems or by different terrestrial systems. In a number of recent FCC decisions, satellite systems and terrestrial systems have been allocated separate frequency bands or separate sections of a band.[5] When both uses have been allowed to operate in the same band, sometimes one is designated as "primary," (i.e. it has first priority with respect to interference) and the other as "secondary," (i.e. it has a lower priority with respect to interference).[6] In some bands, however, satellites and terrestrial users are designated as "co-primary," which means they have equal priority, so generally whichever systems gets built first is protected against interference from the second system. The usual regulatory process is first to allocate spectrum internationally and then domestically to certain types of uses (mobile terrestrial, mobile satellite, fixed terrestrial, fixed satellite, broadcasting (terrestrial), broadcasting satellite, etc.) and sometimes to two or more of those broad classes of service. After this the service or operating rules are developed and finally entities are allowed to apply to obtain licenses.[7] Satellite systems and terrestrial systems have generally been subject to different interference regulations and build out and other service rules for a variety of historical reasons related to such issues as the potential to cause interference and the cost and speed of building systems.[8]

---

[4] See, for example: Gregory L. Rosston and Jeffrey S. Steinberg, "Using Market-Based Spectrum Policy to Promote the Public Interest," <u>Federal Communications Law Journal</u>, vol. 50, no. 1 (December 1997), pp. 87-116; Douglas W. Webbink, "Frequency Spectrum Deregulation Alternatives," FCC Office of Plans and Policy Working Paper No. 2, (1980); Mathtech, Inc., and Telecommunications Systems, <u>Economic Techniques for Spectrum Management: Final Report</u>, by Carson Agnew, Donald A. Dunn, Richard G. Gould and Robert D. Stibolt, a study prepared for the National Telecommunications and Information Administration (December 20, 1979). For a discussion and proposals for ways to increase the efficiency of spectrum use by encouraging secondary markets in the use of the spectrum, see: In the Matter of Principles for Promoting the Efficient Use of Spectrum by Encouraging the Development of Secondary Markets, Policy Statement, 15 FCC Rcd 24,178 (2000). For proposals to develop secondary spectrum markets that mainly emphasize terrestrial wireless use, especially among commercial mobile radio service users, see: In the Matter of Promoting Efficient Use of Spectrum Through Elimination of Barriers to the Development of Secondary Markets, Notice of Proposed Rulemaking in WT Docket No. 00-230, 15 FCC Rcd 24,203 (2000). See also: "Comments of 37 Concerned Economists," in WT Docket No. 00-230, February 7, 2001. For a discussion of how authorization of software defined radios might greatly increase the ease and decrease the costs of implementing more flexible uses of the spectrum, see: In the Matter of Inquiry Regarding Software Defined Radios, Notice of Inquiry in ET Docket No. 00-47, 15 FCC Rcd 5930 (2000). See also: Notice of Proposed Rule Making in ET Docket No. 00-47, 15 FCC Rcd 24,442 (2000).

In the past, writers have argued that government mandated spectrum block allocations often lead to inefficient spectrum uses.[9] Despite increasing amounts of flexibility in allowed spectrum uses, questions are now once again arising about this process of separating spectrum uses and users, because the same block of spectrum, regardless of the purpose for which it is currently allocated and regardless of the current service rules under which licensees may operate, has technical characteristics that make it potentially useful for both satellite and terrestrial use. There are many potential competing demands for the use of any particular block of spectrum. In addition, one aspect of the "convergence" of many different services and technologies offered to consumers and businesses is that there are increasing numbers of situations in which it is likely to be efficient to combine satellite and terrestrial wireless transmission systems in order to deliver particular services to customers, and in some situations it may be efficient to use the same spectrum to operate both satellites and terrestrial facilities.

---

[5]  See, for example: In the Matter of Allocation and Designation of Spectrum for Fixed-Satellite Services in the 37.5-38.5 GHz, 40.5-41.5 GHz, and 48.2-50.2 GHz Frequency Bands; Allocation of Spectrum to Upgrade Fixed and Mobile Allocations in the 40.5-42.5 GHz Frequency Band; Allocation of Spectrum in the 37.0-38.0 GHz and 40.0-40.5 GHz for Government Operations, Report and Order, IB Docket No. 97-95, 13 FCC Rcd 24649 (1998). See also Further Notice of Proposed Rule Making, IB Docket No. 97-95, 66 FR 35,399 (2001).

[6]  See, for example: In the Matter of FWCC Request for Declaratory Ruling on Partial-band Licensing of Earth Stations in the Fixed-Satellite Service That Share Terrestrial Spectrum; and FWCC Petition for Rulemaking to Set Loading Standards for Earth Stations In the Fixed-Satellite Service that Share Terrestrial Spectrum; and Onsat Petition for Declaratory Order that Blanket Licensing Pursuant to Rule 25.111(c) is Available for Very Small Aperture Terminal Satellite Network Operations at C-Band; and Onsat Petition for Waiver of Rule 25.212(d) to the Extent Necessary to Permit Routine Licensing of 3.7 Meter Transmit and Receive Stations at C-Band; and Ex Parte Letter Concerning Deployment of Geostationary Orbit FSS Earth Stations in the Shared Portion of the Ka-band, Notice of Proposed Rulemaking in IB Docket No. 00-203, 15 FCC Rcd 23,127 (2000) ("FWCC NPRM").

[7]  In many recent situations involving the provision of new satellite services or services in new or different frequency spectrum bands, companies have actually filed applications to provide satellite service before the allocations or the service rules were finalized.

[8]  FWCC NPRM, ¶¶ 26-31, 37-44.

[9]  See, for example: Mathtech, Inc., Economic Techniques for Spectrum Management, and Webbink, "Frequency Spectrum Deregulation Alternatives."

Some Further Background Thoughts on Spectrum Allocation, Assignment and Use

Traditionally the rules and regulations governing terrestrial services have been quite distinct from the rules governing satellite services. Among other reasons, this was done because for frequencies higher than several hundred megahertz it was assumed that the coverage area for terrestrial signals, even from tall antenna towers, was limited to relatively small areas. This means that many separate terrestrial systems can transmit on the same frequency in different locations throughout the country without causing "harmful" or "unacceptable" interference. In contrast, satellites located in geostationary orbits often use antennas whose signals cover the whole continental U.S. or at least a large portion of it, unless the satellite uses spot beams to cover a smaller regional area. By locating geostationary satellites at least several degrees apart along the geostationary orbital arc, however, it is possible for highly directional antennas located on earth to pick up the signal from one satellite and reject the signal from another satellite, even if both satellites are transmitting on the same frequency.[10] In addition, the costs of building, launching and operating one geostationary orbit (GSO) satellite will be many times the cost of building and operating one terrestrial facility or even a set of terrestrial facilities covering a large metropolitan market.

When, however, individuals talk about "property rights" in spectrum use, they often assume that it is possible (and perhaps even relatively easy) to define the rights of licensees to transmit a certain kind of signal emission with certain power levels, bandwidth and other characteristics or to be protected to a certain level from interference caused by other licensees. In reality, defining such rights may be extraordinarily complex. Moreover, there is no such thing as a transmitter causing no interference to a receiver operating on a frequency and at a location close to the transmitter. It is only possible to describe an "acceptable" or an "unacceptable" level of interference, or the existence or lack of "harmful" interference, or even measurable interference, defined by such measures as a desired to undesired signal level, or a specific signal strength level or a power flux density or a signal strength level above a certain so called "noise floor." In addition, the amount of "harmful" interference actually received at any location depends not only on the characteristics of the transmitted signal but also on characteristics of the receiving antenna and radio receiver in terms of sensitivity, selectivity, etc. At present, in the U.S., determinations of acceptable or unacceptable interference levels are made through the rule making process at the FCC. An important issue, therefore, is how and by whom are decisions made concerning defining or determining the existence of "harmful" or "unacceptable" interference. This then also raises the issue of who should determine whether additional licensees could share the same spectrum without causing unacceptable interference, or alternatively whether an incumbent licensee should be allowed to prevent possible interfering operations or to accept interference and share the spectrum, presumably in exchange for some level of compensation.

---

[10] In contrast, low earth orbit or medium earth orbit satellites do not remain over one location on the earth to the same extent, which can create much more complicated issues in terms of preventing interference with each other and with other systems, if they are both operating on the same frequencies.

Two Interrelated Issues

The remainder of this paper primarily focuses on two interrelated issues: the first is the degree to which licensees should be given exclusive property rights that allow them, rather than a government agency, to determine when to allow such sharing and under what conditions to allow that sharing, including what kinds of compensation they might accept for that sharing particularly in situations involving satellites and terrestrial users. In other words, the question to be addressed is to what extent should a government agency, e.g. the FCC, make the determination concerning when sharing is or is not feasible or desirable and under what conditions including what kinds of compensation schemes? Related to this is the issue of the extent to which such possible sharing should be analyzed differently depending upon whether the proposed satellite and terrestrial systems are or are not owned and under control of the same party or are owned and under the control of two separate parties. Another way to describe this question is to think of the difference between voluntary sharing by users controlled by the same entity compared to voluntary or involuntary sharing by users controlled by two or more different entities.

The second interrelated issue is a consideration of the impact of one section of the 2000 ORBIT Act, which specified that competing applicants for satellite licenses to provide international or global services may not be selected by auction. In contrast, auctions may be used to select among competing applicants for domestic satellite licenses and many kinds of terrestrial wireless licenses. This anti-auction provision of the ORBIT Act may potentially bias political and regulatory decisions concerning allowable uses of the spectrum. For example, it would not be surprising to see future decisions concerning the allocation of unused spectrum or the reallocation of lightly used spectrum facing substantial pressure to be tilted towards terrestrial users and away from international satellite uses because of the money generating advantage of holding terrestrial auctions.

Property Rights: Company Controlled or Government Controlled?

Within the last several years, a number of proposals to the FCC as well as some decisions made by the FCC raise the question of what degree of exclusive property-like rights should be given to licensees and what property rights should be retained or controlled by the government, e.g. the FCC, and who should have the authority to change the definitions of those rights. Several of the recent proposals involve the provision of terrestrial based wireless services by the same companies which have been licensed, or who have applied for licenses to operate satellite systems. Other proposals relate to the question of whether to allow an independent company to provide terrestrial services in a band of spectrum that has been assigned to or applied for by a satellite service provider. There are also proposals to provide new low power services that may or may not interfere with both satellite and terrestrial services.

For example, satellite digital audio radio service (DARS) licensees Sirius Satellite Radio and XM Satellite Radio have proposed to operate terrestrial repeaters within their own assigned spectrum.[11] Motient Services Inc.[12] and New ICO Global Communications Ltd.[13] have each requested authority to provide ancillary terrestrial services within their satellite spectrum bands.[14] In contrast, Northpoint and MDS America as well as PDC Broadband Corporation and Satellite Receivers, Ltd. have requested authority to provide a terrestrial service within the 12.2-12.7 GHz direct satellite service (DBS) band in which existing DBS licensees (including DirecTV and EchoStar) are currently operating.[15] SkyBridge has also applied to provide non-geostationary satellite service in that same band.

While these particular examples all involve the question of whether the FCC should allow terrestrial wireless systems to operate in spectrum assigned to or applied for by satellite providers, the reverse situation is also likely to arise in the future, i.e. terrestrial licensees or applicants who wish to provide satellite service within their assigned terrestrial bandwidth. For example, TRW requested that the Commission permit fixed satellite service operations in a part of the 39 GHz band designated for terrestrial fixed services. The Commission declined to reserve spectrum solely for satellite use.[16]

---

[11] In the Matter of Establishment of Rules and Policies for the Digital Audio Radio Satellite Service in the 2310-2360 MHz Band, Report and Order, Memorandum Opinion and Order and Further Notice of Proposed Rulemaking in IB Docket No. 95-91 and GEN Docket No. 90-357, 12 FCC Rcd 5754 (1997). ("DARS R&O").

[12] Application by Motient Services Inc. and Mobile Satellite Ventures Subsidiary LLC for Assignment of Licenses and For authority to launch and Operate a Next-Generation Mobile Satellite Service System (March 1, 2001). See also: Public Notice, Report No. SAT-00066 (released March 19, 2001).

[13] Ex parte letter to Chairman Michael K. Powell in IB Docket No. 98-81 (filed March 8, 2001).

[14] In the Matter of Flexibility of Communications by Mobile Satellite Service Providers in the 2 GHz Band, the L-Band and the 1.6/2.4 GHz Band and Amendment of Section 2.106 of the Commission's Rules to Allocate Spectrum at 2 GHz for Use by the Mobile Satellite Service, IB Docket No. 01-185 and ET Docket No. 95-18, FCC 01-225 (released August 17, 2001). ("MSS Flexibility NPRM")

[15] In the Matter of Amendment of Parts 2 and 25 of the Commission's Rules to Permit Operation of NGSO FSS Systems Co-Frequency with GSO and Terrestrial systems in the Ku-Band Frequency Range and amendment of the Commission's Rules to Authorize Subsidiary Terrestrial us of the 12.2-12.7 GHz Band by Direct Broadcast Satellite Licensees and Their Affiliates, Notice of Proposed Rulemaking In ET Docket No. 98-206, 14 FCC Rcd 1131 (1998). ("DBS NPRM"). See also: First Report and Order and Further Notice of Proposed Rule Making, ET Docket No. 98-206, 16 FCC Rcd 4096 (2000). See also: "MDS America Joins Northpoint in Fight for DBS Spectrum," Satellite Week, (May 14, 2001).

[16] See: Amendment of the Commission's Rules Regarding the 37.0-38.6 GHz and 38.6-40.0 GHz Bands, ET Docket No. 95-183, and Implementation of Section 309(j) of the Communications Act – Competitive Bidding, 37.0-38.6 GHz and 38.6-40.0 GHz Bands, PP Docket No. 93-253, Memorandum Opinion and Order, 14 FCC Rcd 12,428 (1999), at ¶¶ 47-49. See also: In the Matter of TRW Inc., Memorandum

However, the commission also stated that "it might be possible and desirable to deploy both terrestrial wireless and satellite facilities."[17] It stated that "satellite operators would be free to provide service in the 39.5-40.0 GHz segment either through a license won at the 39 GHz auction thereby becoming a wireless licensee, or through a post-auction arrangement with a winning bidder."[18]

There are other situations that do not specifically involve satellite v. terrestrial wireless usage of the same spectrum but nevertheless raise similar exclusive property rights issues and questions about the extent to which the government, rather than private parties, should determine the limits of those rights. For example, the authorization of unlicensed ultra-wideband devices to operate at low power in bands allocated for other kinds of uses raises similar questions about who should determine when "undesirable" interference is taking place.[19]

Many of these proceedings are controversial with interested parties presenting quite different and often conflicting points of view. The intent of this paper is not to evaluate the arguments of any particular party in a specific proceeding but rather to consider the underlying policy issues behind the proposals and comments of both supporters and opponents.

The first issue to be discussed concerns who should decide what explicit and what implicit property-like rights are given to licensees to use the spectrum as they see fit. A related issued concerns who should have authority and under what conditions to modify such rules, by, for example, modifying existing allocation rules and then requiring "voluntary" or involuntary relocation of incumbent spectrum users.[20] For example, satellite space station license authorizations or the specific service rules applicable to those authorizations typically specify various characteristics of space station transmission systems such as the number and location of satellites, operating frequencies, power output or power flux density on the earth, limits on out-of-band emissions and the area of coverage of a particular signal strength, etc. Typically such authorizations are silent concerning whether they include the right to transmit signals on the earth on the same

---

Opinion and Order, 16 FCC Rcd 5198 (WTB, 2001) ("TRW MO&O"), ¶9. See also: Order on Reconsideration, IB Docket No. 97-95, 15 FCC Rcd 1766 (1999); Report and Order, IB Docket No. 97-95, 13 FCC Rcd 24,649 (1998).

[17] TRW MO&O, ¶ 3.

[18] TRW MO&O, ¶4.

[19] In the Matter of Revision of Part 15 of the Commission's Rules Regarding Ultra-Wideband Transmission Systems, Notice of Proposed Rule Making in ET Docket No. 98-153, 15 FCC Rcd 12,086 (2000). ("UWB NPRM"). See also: Notice of Inquiry in ET Docket no. 98-153, 63 Fed. Reg. 50184 (September 21, 1998).

[20] For a discussion of the issues concerning incentives to cooperate or oppose relocation of incumbent spectrum users, see: Peter Cramton, Evan Kwerel and John Williams, "Efficient Relocation of Spectrum Incumbents," Journal of Law and Economics, vol. XLI, no. 2 (October 1998), pp. 647-675.

frequencies, even in the areas covered by their signals. Similarly, terrestrial license authorizations or the service rules applicable to those authorizations specify operating frequencies, locations, allowable emission modes and some limit on power levels and out-of-band emissions. Some license authorizations, e.g. TV and radio broadcasting licenses, specify the specific location, power output, antenna height and radiation pattern of the transmitting antenna. Others, such as PCS and cellular license authorizations, specify geographic regions in which transmitters may be placed and also specify the frequencies of operation. The authorizations or service rules also specify limits on power levels, out-of-band emissions and maximum antenna heights. Again, however, such authorizations are usually silent on whether the same frequencies may be used by satellites whose signals cover the same geographic regions. Usually, however, the underlying frequency allocations decisions and the resulting allocations table will indicate whether or not spectrum sharing is allowed, and if so, under what conditions, including what are the rules for sharing or coordination of use.

DARS Terrestrial Repeaters or Gap Fillers

The satellite DARS applicants, XM Radio and Sirius, both proposed to use terrestrial repeaters or gap fillers to improve the quality of service in difficult propagation environments, especially urban areas. They proposed to operate them on the same frequencies as their satellite transmissions and only retransmit those signals.[21] While the Commission approved in principle the use of terrestrial repeaters or gap fillers, it issued a Further Notice of Proposed Rule Making asking how it should be done.[22] A related issue in this proceeding involves the extent to which such DARS terrestrial repeaters or gap fillers may cause interference to the Wireless Communications Service (WCS) licensees who are assigned to frequencies immediately adjacent to and in between the frequencies assigned to the two DARS systems.[23]

ICO and Motient Proposals

In a letter to the FCC, New ICO Global Communications Ltd. stated that adding to its 2 GHz mobile satellite system a terrestrial component that would operate on the same frequencies as its satellite transmissions would allow it to extend service to indoor and urban locations that otherwise would be unserved by a satellite only system and thus would increase the commercial viability of the system.[24] Motient Services Inc. has requested authority to operate terrestrial base stations on frequency bands in which it is authorized to provide satellite service, and to provide coverage in areas where the satellite signal is not sufficiently strong because of foliage or terrain as well as to operate within buildings.[25] A variety of questions have been raised concerning whether to allow such terrestrial service within a satellite band at all, and if so whether only those satellite companies that wish to provide service within the bands for which they hold a satellite license should be allowed to do so or whether other entities should also be allowed to operate terrestrial facilities in those bands.[26] A subsidiary issue involves the question of if entities other than the satellite licensees are allowed to operate terrestrial facilities, should any terrestrial facilities that are authorized be granted through a license auction?[27]

---

[21] DARS R&O, ¶ 140.

[22] DARS R&O ¶ 142.

[23] The extent to which the possibility of interference between DARS terrestrial repeaters and WCS is controversial can be seen by the fact that the FCC Electronic Comment Filing System lists records of over 450 submissions in this docket since 1995.

[24] MSS Flexibility NPRM, ¶¶ 10-11.

Northpoint, SkyBridge etc. DBS Sharing

In the ongoing proceeding concerning possible spectrum sharing in the 12.2-12.7 GHz band between incumbent geostationary DBS licensees, terrestrial applicants such as Northpoint Technology and non-geostationary satellite applicant SkyBridge, there are a variety of issues concerning whether and if so to what extent, the operation of terrestrial service or the operation of non-geostationary satellite service (NGSO) would cause unacceptable or harmful interference to the reception of signals from existing geostationary DBS providers (DirecTV and EchoStar).[28] To the extent that interference may be caused by new users of the spectrum, the overriding policy issue is, once again, whether incumbent licensees or a government agency should decide whether to allow such additional users, and if so, under what conditions.

Ultra-Wideband Transmission Systems

In its recent notice on possible use of unlicensed ultra-wideband systems (UWB), the FCC found that "[w]hile comprehensive tests have not been completed, UWB devices appear to be able to operate on spectrum already occupied by existing radio services without causing interference, which would permit scarce spectrum resources to be used more efficiently."[29] "Part 15 of the Commission's regulations permits the operation of [low power] RF devices without a license from the Commission or the need for frequency coordination. The technical standards contained in Part 15 are designed to ensure that there is a low probability that these devices will cause harmful interference to other users of the radio spectrum."[30] Incumbent licensees question this finding and argue that there

---

[25] MSS Flexibility NPRM, ¶¶ 15-18.

[26] MSS Flexibility NPRM, ¶¶ 23-28.

[27] MSS Flexibility NPRM, ¶¶ 37-40.

[28] DBS NPRM, ¶¶ 2-8. See also: Analysis of Potential MVDDS Interference to DBS in the 12.2-12.7 GHz Band, Mitre Technical Report MTR 01W0000024 (the Mitre Corporation, April, 2001). The level of controversy in this proceeding is indicated by the fact that the FCC Electronic Comment Filing system lists records of over 845 submissions in this docket since 1998.

[29] UWB NPRM, ¶ 1.

[30] UWB NPRM, ¶ 2. Specifically, "operation of an intentional, unintentional or incidental radiator is subject to the conditions that no harmful interference is caused and that interference must be accepted that

is the possibility of substantial interference.[31] Thus the policy issue is, once again, who should determine the existence of acceptable or unacceptable levels of interference and what those levels may be: spectrum licensees or a government regulatory agency.

Who Determines and Who Can Modify Property Rights?

The question to be considered with respect to all of these specific examples and many others likely to arise in the future, is not whether some level of interference might be caused and if so, specifically how it might be mitigated. Nor is the basic issue who should be responsible for what kinds of interference mitigation if such mitigation is necessary. Rather the basic issue is more fundamental than that: it is whether a government agency or actual or potential service providers should decide whether or not they believe there will or will not be harmful interference. In either case, the issue then moves to the question of who has the right to decide whether to allow companies to operate that may or may not cause interference to other companies and how much interference to allow.

One argument that has been made is that in order for the spectrum to be used efficiently, a government agency needs to decide how much interference is significant or harmful and if such interference does exist, who has the responsibility to mitigate it. In essence, according to this view, the government would decide or determine the initial property rights and also have the ability to revise or modify those rights after following the proper administrative procedures.

However, advocates of giving spectrum users more property-like rights argue that companies rather than a government agency should generally make these decisions. In particular, government determined property rights may be either too restrictive or not restrictive enough. That is, the government may restrict the ability of a party to provide services within its authorized band (e.g., by preventing a satellite from providing terrestrial service or a terrestrial licensee from providing satellite service in its band even when it could be done without causing interference to others). On the other hand, the government may allow independent companies to operate within a band after determining that significant interference will not be caused to incumbents or even after deciding that some significant level of interference will be caused to incumbents, even without the agreement of incumbents.

There is an additional and related policy question. If a new entrant will cause at least noticeable interference (however determined) to incumbent licensees, but will it also increase the level of competition in the provision of some set of services that are provided

---

may be caused by the operation of authorized radio stations, by another intentional or unintentional radiator, by industrial, scientific and medical (ISM) equipment, or by an incidental radiator." 47 CFR § 155 (b) (2000).

[31] See: Comments Requested on Test Data Submitted by the National Telecommunications and Information Administration Regarding Potential Interference from Ultra-Wideband Transmission, Public Notice, DA-01-171 (January 24, 2001).

either by the incumbent spectrum users or by providers of totally different services in different frequency bands or using other media such as wire, coaxial cable or fiber optic cable, to what extent should the government balance the increased level of competition against the increase in the level of interference to incumbents? To the extent that incumbents attempt to prevent new entry and sharing, how does one determine whether such behavior is caused by genuine fear of interference, by a desire to prevent entry of potential competitors even if no interference will be caused, or by both?

An argument against allowing involuntary sharing between satellite and terrestrial users is that companies may have little incentive to resolve real or hypothetical interference problems when such involuntary sharing is mandated by the government. In contrast, when the same party operates both satellite and terrestrial services, it is much more likely to make efficient decisions about dealing with real or hypothetical interference. Or, in the jargon of economists, when two parties are involved they may find it difficult to deal with potential externalities, whereas if the same company provides both services, it may find it much easier to internalize any potential externalities. When one company provides both services either as complements or substitutes, its incentives will be aligned to minimize interference or at least to establish an efficient level of interference .

Spectrum or License Auctions

In the U.S., selecting among mutually exclusive applications for spectrum allocated for terrestrial use may be done by auction (with some restrictions), whereas spectrum used for satellites, if it might be used for international or global use, can not be auctioned. Although the primary policy goal of auctions should not be to raise money for the U.S. treasury, but instead to lead to the best and fastest use of the spectrum, the revenue raising aspects of auctions are a powerful political force.[32] How does one weigh the possible benefits of allowing flexibility in terrestrial operations in the satellite band against the possibility that requests for authorization of terrestrial uses will be used as a way to avoid bidding in an auction?

Many papers have been written concerning the benefits of using auctions to chose among competing spectrum applicants.[33] Proponents of the use of such license auctions assert

---

[32] For an example of a story that suggests the arguments used in favor of auctions and against issuing licenses without using an auction, see: Andrew Backover and Paul Davidson, "8 Companies Get Free Spectrum Licenses: Irked Wireless Firms Say They Would Pay Billions," USA Today, (July 17, 2001), p. 57.

[33] See, for example, Evan Kwerel and Walter Strack, "Auction Spectrum Rights," U.S. Federal Communications Commission (February 20, 2001); Martin Cave and Tommaso Valletti, "Are Spectrum Auctions Ruining Our Grandchildren's Future," Info, vol. 2, no. 4 (August 2000), pp. 347-350; Peter Cramton, "The Efficiency of the FCC Spectrum Auctions," The Journal of Law and Economics, vol. XLI, no. 2, (October 1998), pp. 727-736; Thomas W. Hazlett, "Assigning Property Rights to Radio Spectrum Users: Why Did FCC Licenses Auctions Take 67 Years?", The Journal of Law and Economics, vol. XLI no. 2, (October 1998), pp. 529-575; William Kummel, "Spectrum Bids, Bets, and Budgets: Seeking an Optimal Allocation and Assignment Process for Domestic Commercial Electromagnetic Spectrum Products, Services, and Technology," Federal Communications Law Journal, vol. 48, no. 3, (June, 1996), pp. 511-544; Evan Kwerel and Alex D. Felker, "Using Auctions to Select FCC Licensees," FCC OPP

that auctions lead to faster grant of licenses than any of the regulatory alternatives such as lotteries or comparative hearings or so-called "beauty contests." Because the winner of the auction is the company that bids the most in the auction, auctions generally allow the entities who value the license most to obtain the license. Presumably, therefore, the winner is generally the entity best able efficiently to provide service desired by consumers and businesses. Moreover, auctions provide information which indicates how much bidders believe the spectrum is worth, and therefore provide an indication of the opportunity cost of using or not using the spectrum.[34] Many auction proponents also suggest that the primary goal of such auctions should not be to raise revenue for the government but only to lead to rapid and efficient use of the spectrum.[35] In addition to the U.S., several foreign countries such as Brazil, Mexico and Australia have held auctions or plan to hold auctions for domestic satellite licenses.[36]

At the same time, however, a number of arguments have been raised against using spectrum auctions, especially in the international context.[37] For example, it has been argued that when satellite signals cover multiple countries, if there would be multiple sequential auctions, it would be difficult for an applicant to decide how much to pay for spectrum rights in the first country, assuming that the applicant needed to bid for rights to obtain access to the other countries. Hence there would be substantial uncertainty and it would be difficult to develop rational business plans. In particular, the applicant might fear that there would be a holdout problem, i.e. the last country might attempt to extract

---

Working Paper No. 16 (May 1985). See also: Lawrence D. Roberts, "A Lost Connection: Geostationary Satellite Networks and the International Telecommunication Union," Berkeley Technology Law Journal, vol. 15, (2000), pp. 1095-114; and Martin A. Rothblatt, "An ITU Stock Market in Orbital Slots Practice Run in the ISY," paper presented at the 42nd Congress of the International Astronautical Federation (October 5-11, 1991).

[34] Of course, if the class of entities allowed to bid in the auction is restricted, or if the auction winner is already an incumbent monopolist or can use the newly obtained spectrum to become a monopolist, then the amount bid in the auction will not necessarily reflect the competitive market price for using the spectrum or the opportunity cost of its use.

[35] Unfortunately, however, as Eli Noam puts it, "Auctions Inevitably Deteriorate into Revenue Tools." Eli Noam, op. cit., pp. 772-775.

[36] "Loral Skynet Wins Brazilian Slot Auctions," Satellite Today (March 18, 1999); "Mexican Government Poised to Auction Satellite Orbital Location License," Corporate Mexico: Reforma (Mexico), (July 12, 2001); "Two Satellite Slots Available for Australian Broadcasters," Asia Pulse (Mary 7, 2001).

[37] See, for example, Clayton Mowry, "Auctions? No, no, no." Satellite Communications (February 1, 1997); Louis Jacobson, "Lobbying & Law: Washington's Rocket Man," National Journal (June 27, 1998); Curt Harler, "Intelsat goes into Orbit," Communications International (April 1, 2000); "Regulatory Review: No Satellite Auctions," Via Satellite (June 14, 2000); Charles L. Jackson, John Haring, Harry M. Shooshan III, Jeffrey H. Rohlfs and Kirsten M. Pehrsson, "Policy Harms Unique to Satellite Spectrum Auctions," Strategic Policy Research, Inc., A Study Prepared for the Satellite Industry Association (March 18, 1996). See also: Harold Gruber, "Spectrum Limits and Competition in Mobile Markets: The Role of License Fees," European Investment Bank, (June 2000). See also: Eli Noam, "Spectrum Auctions: Yesterday's Heresy, Today's Orthodoxy, Tomorrow's Anachronism. Taking the Next Step to Open Spectrum Access," The Journal of law and Economics, vol. XLI. no. 2 (October 1998), pp. 765-790.

all remaining economic rents from the applicant or may even, for some political reason, set a minimum price so high that the applicant would be unable or unwilling to pay for it. Indeed, the last country may set the minimum price in order to restrict entry, to protect an incumbent service, or even deny service so that no applicant can provide such service. While these potential problems concerning sequential international auctions are real and should not be ignored, they do not lead to an inevitable conclusion that auctions should be forbidden.[38]

The ORBIT Act's Anti-Auction Provision

Since 1993 the FCC has had authority to chose among competing license applicants for many services using an auction. Indeed, between 1994 and 2000, the FCC has held at least 39 auctions, awarded at least 15,304 licenses through auction and the net amount bid in those auctions has totaled about $41.6 billion.[39] By statute, certain classes of license applications such as public safety and amateur radio are exempt from auctions. In addition, the Open-market Reorganization for the Betterment of International Telecommunications Act, the ORBIT Act, passed in 2000, states in relevant part:

> Not withstanding any other provision of law, the Commission shall not have the authority to assign by competitive bidding orbital locations or spectrum used for the provision of international or global satellite communications services. The President shall oppose in the International Telecommunication Union and in other bilateral and multilateral fora any assignment by competitive bidding of orbital locations or spectrum used for the provision of such services.[40]

It is not entirely clear precisely what is covered by this provision and when it applies and when it does not apply. Most satellite systems proposed recently are designed to provide international or global coverage or usage. Others may be designed only to serve the U.S. market. Apparently the former class of applications would be covered by the ORBIT Act anti-auction provisions and the latter would not. However, it is not clear how this law would apply if an applicant changed his or her plans and also how applications should be treated if some applicants apply to provide only domestic service and others apply to provide international service in the same band.

---

[38] It is hypothetically possible that absent an agreement to forbid auctions, transactions costs would become so high that no agreement could be reached, but there should be alternatives to deal with potential hold-out problems that do not require forbidding auctions.

[39] "FCC Auction Summary," as of 6/15/01 at http://www.fcc.gov/wtb/auctions/welcome/html. Of course, as the ongoing controversy concerning the licenses originally won by NextWave before it went bankrupt demonstrate, the government has not actually collected all of the $41.6 billion.

[40] Communications Satellite Act of 1962, Section 647. (Provision added by the ORBIT Act, Public Law 106-180, 114 Stat. 48 (2000)).

It is also not entirely clear how this provision came to be added to the ORBIT Act, even though the U.S. satellite industry supported its inclusion, because most of the Act's provisions relate specifically to INTELSAT and INMARSAT. What is clear, however, is that many members of the U.S. satellite industry believe that this anti-auction provision helps them, because they will not be required to bid in a U.S. license auction to obtain additional satellite orbital positions or spectrum.[41] Unfortunately, the industry members are very likely wrong in this belief about how the ORBIT Act will help them because they have not fully evaluated the possibilities of alternative non-satellite uses of the same spectrum nor the political importance of the revenue to be obtained for the U.S. Treasury from license auctions. In fact, this provision may lead to strong political pressures to reallocate more spectrum to auctionable terrestrial users and less spectrum to non-auctionable satellite users. Thus, while satellite industry members may believe that they are protected by this statutory provision, it can be argued that they would be far better off without it.

Conclusions

There are two major implications suggested by this paper. First, wherever feasible, it would be desirable to give companies the maximum exclusive property-like rights.[42] In other worlds, it would be desirable to allow them and not a government agency to decide if sharing should be allowed and if so, under what conditions. Ideally, government agencies should only become involved in such decisions if there is a problem concerning monopoly control of the spectrum by incumbents or potential applicants.[43] A corollary to this conclusion is that for companies making business decisions, changing rules such as the property rights implicitly attached to licenses, may impose substantial uncertainty and thus real economic costs on such firms.[44]

Second, with respect to auctions and the ORBIT Act, that particular provision of the law should be repealed. Although there may be situations in which the use of auctions is undesirable, a flat out prohibition on auctions involving international or global services will surely lead to undesirable results. Over the long run that anti-auction provision will likely harm satellite companies and their ability to provide service to customers, far more than it will help satellite companies and their customers.[45]

---

[41] See, for example: Clayton Mowry, "Auctions? No, no, no." Satellite Communications (February 1, 1997); Louis Jacobson, "Lobbying & Law: Washington's Rocket Man," National Journal (June 27, 1998); Curt Harler, "Intelsat goes into Orbit," Communications International (April 1, 2000); Regulatory Review: No Satellite Auctions," Via Satellite (June 14, 2000).

[42] An important assumption is that the benefits of added flexibility will exceed any potential transactions costs or externality problems that might arise from that added flexibility.

[43] For a similar view, see: "Comments of 37 Concerned Economists" in WT Docket No. 00-230 (February 7, 2001).

[44] Of course, if the changes give firms far more flexibility in future activities, even if those changes introduce uncertainty, the benefits of the changes may still outweigh the costs. On the other hand, if changes force incumbent firms to vacate existing spectrum or to adopt new more technically efficient (but not necessarily more economically efficient) spectrum techniques, those changes will almost surely impose costs on those firms even though the changes might also benefit new potential entrants. In addition, it is possible that additional flexibility might make it more difficult for the government or for competitors to monitor the activities of incumbent companies, including determining who is causing interference.

If the anti-auction prohibition on satellites remains, public policy decisions on the best uses of the spectrum should not be based upon whether use of the spectrum will generate more or less auction revenue for the U.S. treasury. Auctions can provide important and useful price signals, but only if many competing potential users can bid in an auction. Price signals are likely to be highly misleading if certain classes of users are excluded from bidding. And, in particular, policy decisions concerning the "best" use of the spectrum should not be based on whether or not the licensee bid in an auction, when certain classes are forbidden from bidding.

---

[45] Of course, many satellite industry members still believe that auctions for the right to provide international satellite services could lead to higher costs of operations and holdout problems with some countries.